\documentclass[sigconf,screen]{acmart}

\usepackage[nolist]{acronym}
\usepackage{xspace}

\begin{acronym}
    \acro{LLM}{Large Language Model}
    \acro{SE}{Security-related Element}
    \acro{EV}{Electric Vehicle}
    \acro{SWATTR}{SoftWare Architecture Text Trace link Recovery}
    \acro{SecDReqAn}{Security- and Design-related Requirements Analyses}
    \acro{ArDoCo}{Architecture Documentation Consistency}
    \acro{ArCoTL}{ARchitecture-to-COde Trace Linking}
    \acro{TransArC}{Transitive links for Architecture and Code}
    \acro{SAD}{software architecture documentation}
    \acro{NLP}{natural language processing}
    \acro{TLR}{traceability link recovery}
    \acro{ID}{inconsistency detection}
    \acro{RI}{recommended instance}
    \acro{SAM}{software architecture model}
    \acro{pp}{percentage points}
    \acro{UML}{Unified Modeling Language}
    \acro{PCM}{Palladio Component Model}
    \acro{UME}{unmentioned model element}
    \acro{MME}{missing model element}
    \acro{NLSAD}{natural language software architecture documentation}
    \acro{MAP}{mean average precision}
    \acro{JSD}{probabilistic Jensen Shannon model divergence}
    \acro{VSM}{vector space model}
    \acro{LSI}{latent semantic indexing}
    \acro{WMD}{Word Mover's Distance}
    \acro{IR}{information retrieval}
    \acro{ML}{machine learning}
    \acro{DFD}{data flow diagram}
    \acro{CoT}{chain of thought}
\end{acronym}

\newcommand{\EV}{\ac{EV}\xspace}

\newcommand{\SEFFs}{service effect specifications\xspace}
\newcommand{\SEFF}{service effect specification\xspace}

\usepackage[english]{babel}
\usepackage{csquotes}

\usepackage{tabularx}
\newcolumntype{Y}{>{\raggedleft\arraybackslash}X}
\newcolumntype{Z}{>{\centering\let\newline\\\arraybackslash\hspace{0pt}}X}

\usepackage{adjustbox}

\usepackage[dvipsnames]{xcolor}

\setcopyright{cc}
\setcctype{by}
\copyrightyear{2026}
\acmYear{2026}
\acmDOI{XXXXXXX.XXXXXXX}
\acmConference[ASE '26]{the 41st IEEE/ACM International Conference on Automated Software Engineering}{October 12--16,
  2026}{Munich, Germany}
\acmISBN{978-1-4503-XXXX-X/2026/10}

\usepackage{pifont}
\newcommand{\cmark}{\ding{51}}%

\usepackage{makecell}
\usepackage[inline]{enumitem}

\newsavebox\CBox
\def\textBF#1{\sbox\CBox{#1}\resizebox{\wd\CBox}{\ht\CBox}{\textbf{#1}}}

\begin{document}

\title{The EVerest Dataset for Secure Software Engineering}


\author{Sophie Corallo}
\orcid{0000-0002-1531-2977}
\email{sophie.corallo@kit.edu}
\author{Debora Grupp}
\orcid{0009-0000-1655-8226}
\email{udimk@student.kit.edu}
\affiliation{%
  \institution{Karlsruhe Institute of Technology}
  \city{Karlsruhe}
  \country{Germany}
}

\author{Dominik Fuchß}
\orcid{0000-0001-6410-6769}
\email{dominik.fuchss@kit.edu}
\author{Jan Keim}
\orcid{0000-0002-8899-7081}
\email{jan.keim@kit.edu}
\affiliation{%
  \institution{Karlsruhe Institute of Technology}
  \city{Karlsruhe}
  \country{Germany}
}

\author{Frederik Reiche}
\orcid{0000-0002-5993-0558}
\email{frederik.reiche@kit.edu}
\author{Tobias Hey}
\orcid{0000-0003-0381-1020}
\email{hey@kit.edu}
\author{Anne Koziolek}
\orcid{0000-0002-1593-3394}
\email{koziolek@kit.edu}
\affiliation{%
  \institution{Karlsruhe Institute of Technology}
  \city{Karlsruhe}
  \country{Germany}
}

\renewcommand{\shortauthors}{Corallo et al.}

\begin{abstract}
End-to-end security verification, from requirements through architecture to code, requires datasets that span all three artifact types with fine-grained security labels.
No existing dataset provides this combination.
We present the \emph{EVerest dataset}, a multi-artifact resource based on EVerest, an industry-driven open-source software stack for electric vehicle charging stations.
The dataset includes 84 manually elicited security requirements annotated with security objectives, 1{,}445 fine-grained security elements (components, entities, data, data flows, states, etc.), acceptance windows, coreferences, and architectural trace links, as well as the EVerest software architecture model, source code, and natural language documentation.
It enables research on security requirements classification, named entity recognition, architectural trace linking, and design-time or code-level security verification.
During dataset creation, a real security weakness (CWE-1295) was identified, disclosed to the project maintainers, and subsequently fixed.
The dataset is publicly available \cite{EVerest_dataset_2026}.
A short video is available at \url{https://youtu.be/pnn1uqpomvQ}.
\end{abstract}




\maketitle

\section{Introduction}

Ensuring software security requires holistic verification across the development chain, from natural language requirements through software architecture to implementation.
This includes (i)~classifying security-relevant requirements and their objectives; (ii)~recognizing security-relevant named entities in requirements and tracing them to architectural elements; and (iii)~checking whether the implementation satisfies the resulting constraints.
Omitting any of these steps can introduce security weaknesses.
Though each step is well-studied, their combination is rarely explored due to missing datasets. 
Existing datasets either cover only requirements (e.g., PROMISE~NFR) or include code but no architecture (e.g., DiverseVul).
No dataset spans requirements, architecture, \emph{and} code, preventing research on end-to-end security verification.

\begin{figure}
    \centering
    \includegraphics[width=\linewidth]{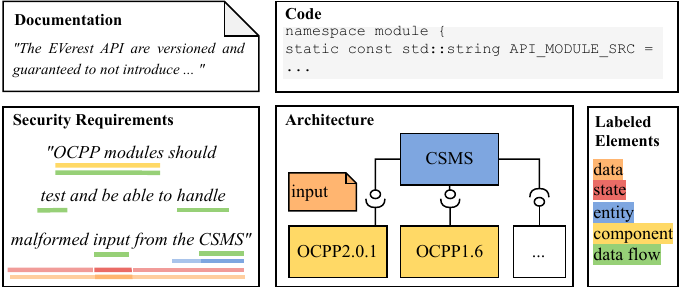}
    \caption{EVerest dataset: documentation, labeled security requirements, architecture, source code, and arch. trace links.}
    \Description[Overview of the artifacts in the EVerest dataset]{The EVerest dataset includes labeled security requirements, architecture, code, and architectural trace links. The requirement in the figure is exemplary labeled with: data, state, entity, component, and data flow. The labeled entity demonstrates the labeling of an acceptance window: \enquote{the CSMS} is the maximal acceptable phrase, whereas \enquote{CSMS} should always be contained.}
    \label{fig:articts_overview}
\end{figure}

We present the \emph{EVerest dataset}, a multi-artifact resource derived from EVerest, an industry-driven open-source software stack for \EV charging stations.
As shown in \autoref{fig:articts_overview}, it contains
\begin{enumerate*}
    \item \textbf{84 security requirements} manually elicited from EVerest documentation and developer interviews, annotated with security objectives, 1{,}445 fine-grained security elements (components, entities, data, data flows, states, etc.), acceptance windows, references, and coreferences;
    \item a \textbf{software architecture} model with requirement-to-architecture trace links;
    \item the \textbf{source code} of EVerest; and
    \item its \textbf{natural language documentation}.
\end{enumerate*}
\noindent
Available under the Apache License~2.0~\cite{EVerest_dataset_2026}, the dataset supports research on security requirements classification, named entity recognition, architectural trace linking, and design-time or code-level security verification.
We also uncovered a real security weakness: a violation of the authentication token storage requirement (CWE-1295).

\section{Related Datasets}
To contextualize the EVerest dataset, we survey requirements engineering and software security datasets. We focus on datasets with textual artifacts and compare them by artifact coverage, security-label granularity, and availability. \autoref{tab:rw-Datasets} summarizes the results.

Dalpiaz~\cite{dalpiaz_userstories_2018} provides requirements from 22 web sources as user stories. QuRE~\cite{QuRE_Dataset_2025} contains Mercedes-Benz specification requirements annotated with weak words and defects. PURE~\cite{PURE_Paper_2017} collects requirements automatically extracted from 79 online documents and labels structural properties. Later studies re-labeled PURE for other tasks, such as distinguishing requirements from non-requirements~\cite{ReqExp_2021}. However, some variants were not published~\cite{pure_ext_not_published, pure_ext_not_published_2} or lack sufficient description~\cite{pure_nfr_ext}. 
PROMISE NFR~\cite{PROMISE_2007} is a classification dataset that contains requirements from 15 projects, labeled with one functional and eleven non-functional classes, two of which are security-relevant. The dataset was further relabeled and extended by Dalpiaz et al.~\cite{dalpiaz_PROMISEextension_2019} and in NICE~\cite{NICE_Dataset_2025}.

Other datasets better support security requirements classification. Slankas and Williams~\cite{slankas_williams_2019} labeled collected requirements with functional and 14 non-functional classes, including six security-related ones.
Riaz et al.~\cite{riaz_original_paper_2014} explicitly label requirements by security objectives. Varenov et al.~\cite{varenov_2024} similarly provide security requirements from other datasets for multi-class labeling.

All datasets above are limited to requirements and lack complementary artifacts such as architecture or source code. Wang et al.~\cite{OSS-Security-Req_2019} go further by linking security- and non-security-labeled requirements to source code. SecReq~\cite{SecReq_2021} labels requirements from three public specifications, whose design or architecture can partly be inferred; however, the exact source documents are undisclosed.

Security datasets with multiple artifacts usually exclude requirements. BADS~\cite{BADS_arxiv_2025} pairs code snippets with natural-language descriptions of intent, vulnerabilities, and severity. DiverseVul~\cite{DiverseVul_arxiv_2023} provides vulnerability-related functions from security commits, annotated with commit messages and CWE identifiers.

No existing dataset covers requirements, architecture, and code while providing labels for both security-objective and generic requirements classification. EVerest addresses this gap.

\section{About EVerest}

EVerest is an open-source, modular software stack for \EV charging stations, covering the full range from low-level hardware drivers to high-level charging protocols.
Initiated by PIONIX GmbH and hosted by the Linux Foundation Energy, it is developed openly on GitHub~\cite{everestGithub} with an active community that meets regularly.
\textit{everest-core}, containing the central charge controller logic, was first released in December~2022.
By June~2024, roughly 40 contributors had grown the codebase to approximately 50~kloc across more than 500 files, written in C++, C, JavaScript, Python, and Rust.
While the project ships user-facing documentation, it provides neither formal requirements documents nor architectural models.

\begin{table}
\caption{Related datasets by artifact type. Parentheses denote partial coverage (\cmark) or derivable artifacts (-).}
\setlength\tabcolsep{2pt}
\renewcommand{\arraystretch}{0.9}
\small
\begin{tabularx}{\columnwidth}{lZZZZZZcZZZZZZ}
\toprule
\textBF{Artifact} & \rotatebox{90}{\small{\textBF{Dalpiaz}~\cite{dalpiaz_userstories_2018}}} & \rotatebox{90}{\small{\textBF{QuRE}~\cite{QuRE_Dataset_2025}}} & \rotatebox{90}{\small{\textBF{PURE}~\cite{PURE_Paper_2017}}} & \rotatebox{90}{\small{\textBF{PROMISE NFR}~\cite{PROMISE_2007}}} & \rotatebox{90}{\small{\textBF{Dalpiaz et al.}~\cite{dalpiaz_PROMISEextension_2019}}} & \rotatebox{90}{\small{\textBF{NICE}~\cite{NICE_Dataset_2025}}} & \rotatebox{90}{\shortstack{\small{\textBF{Slankas \&}}\\\small{\textBF{Williams}~\cite{slankas_williams_2019}}}} & \rotatebox{90}{\small{\textBF{Riaz et al.}~\cite{riaz_original_paper_2014}}} & \rotatebox{90}{\small{\textBF{Varenov et al.}~\cite{varenov_2024}}} & \rotatebox{90}{\small{\textBF{Wang et al.}~\cite{OSS-Security-Req_2019}}} & \rotatebox{90}{\small{\textBF{SecReq}~\cite{SecReq_2021}}} & \rotatebox{90}{\small{\textBF{BADS}~\cite{BADS_arxiv_2025}}} & \rotatebox{90}{\small{\textBF{DiverseVul}~\cite{DiverseVul_arxiv_2023}}} \\
\midrule
\textBF{Req.}     & \cmark                                                               & \cmark                                                   & \cmark                                                   & \cmark                                                       & \cmark                                                                           & \cmark                                                   & \cmark                                                                        & \cmark                                                                   & \cmark                                                          & \cmark                                                                & \cmark                                                 & (\cmark)                                                 & (\cmark)                                                             \\
\textBF{F/Q}      & -                                                                    & -                                                        & (-)                                                      & \cmark                                                       & \cmark                                                                           & \cmark                                                   & \cmark                                                                        & -                                                                        & -                                                               & -                                                                     & -                                                      & -                                                        & -                                                                    \\
\textBF{Sec.}     & -                                                                    & -                                                        & (-)                                                      & \cmark                                                       & -                                                                                & (\cmark)                                                 & (\cmark)                                                                      & \cmark                                                                   & -                                                               & \cmark                                                                & \cmark                                                 & -                                                        & -                                                                    \\
\textBF{Sec.Obj.} & -                                                                    & -                                                        & -                                                        & -                                                            & -                                                                                & -                                                        & (\cmark)                                                                      & \cmark                                                                   & \cmark                                                          & -                                                                     & -
& -                                                        & -                                                                    \\
\textBF{Arch.}    & -                                                                    & -                                                        & -                                                        & -                                                            & -                                                                                & -                                                        & -                                                                             & -                                                                        & -                                                               & -                                                                     & (-)                                                    & -                                                        & -                                                                    \\
\textBF{Code}     & -                                                                    & -                                                        & -                                                        & -                                                            & -                                                                                & -                                                        & -                                                                             & -                                                                        & -                                                               & (-)                                                                   & -                                                      & \cmark                                                   & \cmark                                                               \\
\bottomrule
\end{tabularx}
\label{tab:rw-Datasets}
\end{table}

\section{Dataset Construction}

Since EVerest lacked both documented security requirements and a formal software architecture model, we constructed the dataset from scratch in cooperation with PIONIX in four steps:
\begin{enumerate*}
    \item We elicited coarse-grained security requirements from the EVerest community via an online questionnaire.
    \item We conducted semi-structured interviews with EVerest developers to refine these requirements to the architectural level, yielding fine-grained requirements that explicitly reference specific EVerest components.
    \item We derived a software architecture model from the EVerest source code.
    \item Multiple annotators labeled acceptance windows, references, and coreferences of security elements in the requirements, as well as the contained trace links; disagreements were resolved through inter-annotator agreement.
\end{enumerate*}

To broaden the dataset's applicability, for example, for coarse-grained requirements classification, we additionally included architecture excerpts from the EVerest project documentation.

\subsection{Questionnaire}

Employing multiple elicitation techniques has been shown to improve the completeness and quality of requirements~\cite{Yosuf2015}.
Thus, we designed an online questionnaire for step~(1) to systematically elicit initial security requirements from the broader EVerest community.
Given that participants may not have formal security expertise, we structured the questionnaire around four established security objectives (confidentiality, integrity, availability, and authentication).

The questionnaire comprises seven parts: an introduction, a demographics section, four objective-specific sections, and a closing section for uncategorized requirements.
The introduction outlines the structure, defines security requirements, and provides writing guidelines.
The demographics section captures participants' professional background and their self-assessed familiarity with EVerest and software security proficiency.
For each security objective, participants are provided with the corresponding ISO~27000 definition~\cite{iso2700:2018}, domain-specific elicitation prompts~\cite{miller_2009}, and an example to anchor their responses.

Prior to distribution, we conducted a pilot study with two doctoral researchers with a security background, leading to minor refinements of the instrument.
The questionnaire was then disseminated via the EVerest developer mailing list and announced at two consecutive weekly community meetings.
Seven participants responded: three from Pionix and four from other organizations.

In total, participants submitted 67 security requirements.
To ensure data quality, we applied a systematic cleanup procedure: removing responses that were overly general or unrelated to security, resolving coreferences, discarding non-requirement sentences, excluding answers that merely restated questionnaire examples, and correcting category misclassifications.
After cleanup, 57 requirements were retained, 10 on authentication, 25 on confidentiality, 8 on integrity, 11 on availability, and 5 uncategorized, with two requirements carrying multiple labels.
The left side of \autoref{fig:elicitation} shows a representative requirement obtained from the questionnaire.

\subsection{Interviews}

\begin{figure}
    \centering
    \includegraphics[width=.99\linewidth]{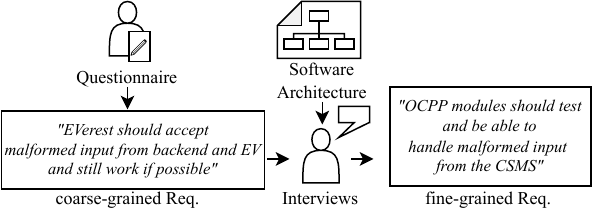}
    \Description[Requirements Elicitation]{From a questionnaire, coarse-grained security requirements were retrieved. In interviews with developers, these were refined using the architecture.}
    \caption{Req. Elicitation: Coarse-grained requirements are retrieved via a questionnaire and refined in interviews.
    }
    \label{fig:elicitation}
\end{figure}

As the questionnaire requirements were high-level and did not refer to EVerest components, we conducted semi-structured interviews with four EVerest developers in step~(2) to refine the coarse-grained requirements to the architectural level (as depicted in \autoref{fig:elicitation}).

Each interview consisted of three parts.
First, the interviewer gave a brief introduction, shared the study goals and consent information, and collected general background information from the participant.
In the main part, the interviewer presented a coarse architecture diagram of EVerest and, using two examples in a shared interview-specific document, explained how coarse-grained requirements should be refined by explicitly naming responsible EVerest components.
The document also contained a selection of 30 security requirements from the questionnaire.
The interviewee was then asked to think aloud and write down fine-grained specifications for the given requirements; the interviewer supported them with writing-recommendation hints and reminders to reference specific components.
Finally, where time constraints led to incomplete sentences, the interviewer revised them afterward and sent the result to the interviewee for confirmation or correction.

After a pilot with four doctoral researchers introduced to EVerest, we remotely interviewed four EVerest developers: three from Pionix and one from Chargebyte. Each interview lasted about 90 minutes.

In total, we retrieved 93 fine-grained requirements: 41 on confidentiality, 14 on integrity, 18 on availability, 10 on authentication, and 10 other security requirements.

\subsection{Architecture Modeling}

EVerest is built around loosely coupled modules communicating exclusively via MQTT; the framework manages their instantiation, communication, and dependency resolution.
We therefore focused on the EVerest core, which encapsulates the central application logic and orchestrates the surrounding modules.

The architecture model was constructed as a Palladio component model~\cite{palladio} as part of a practical course in a computer science master's program, in which three students developed it under weekly supervision by three doctoral researchers in software engineering from the EVerest repository as of 3rd June 2024 (commit 177a8e6).

Each module was represented as a component with interface descriptions capturing the exchanged MQTT messages, including external modules interacting with internal ones.
As the documentation lacked sufficient detail, all modeling decisions were grounded in the source code.
For every contained method, the students created a \SEFF.
\SEFFs describe the internal behavior of a component’s service, capturing its control flow, resource demands, and interactions with other services.
The resulting model comprises 29 components, 34 interfaces, and 144 \SEFFs.
In the assembly model, each component is represented with its required and provided interfaces; the deployment scenario assumes a single charging station alongside external entities such as an update server and a charging station management system, though alternative deployments can be derived straightforwardly.
Finally, the Palladio usage model captures 14 scenarios extracted from the existing documentation and source code, covering, among others, firmware updates, charger enable/disable operations, and limit configuration.

\subsection{Labeling Requirements}

To ensure that the elicited requirements were actually requirements, two annotators labeled the texts as requirements or not requirements. 
As a result, 84 security requirements remained.

The labels of the security objectives of the security requirements stem from the elicitation.

The labeling of security elements was performed by three annotators (one PhD researcher, two students).
As element types, classes of SecLan~\cite{SecLan} are used, which model common concepts in design-level security analyses/specifications and in security checks at the implementation level.
The concepts, derived from an exploratory study, encompass and generalize all security elements of interest for security analyses.
These include components, entities, data, states, hardware nodes, hardware connections, control flows, data flows, and internal activities.
For this, the annotators initially started with the original definitions.
Due to uncertainties, they reworked the definitions and added examples to reach a shared understanding.
The resulting definitions and examples are documented in the dataset.

In shared annotation sessions, elements were labeled at two granularities.
For explicitly mentioned components, entities, data, and nodes (e.g., \enquote{payment provider}), annotators marked acceptance windows: the \emph{short sequence} captures the minimal text identifying the element, while the \emph{long sequence} additionally encompasses articles, determiners, and modifying adjectives and adverbs (e.g., \enquote{CSMS} vs.\ \enquote{the CSMS}, as shown in \autoref{fig:articts_overview}).
Elements not unambiguously referring to an architectural element (e.g., \enquote{the provided interface}) were marked as \emph{references}; \emph{coreferences} point back to the short sequence of the referenced element.
For connections and data flows, annotators marked and linked the \emph{source} and \emph{target} elements; additional type-specific attributes were recorded where applicable, for instance, data flow labels capture the \emph{data} transmitted and the \emph{transmission} verb.
All label definitions, descriptions, and supporting references are documented in the dataset's labeling guidelines.

In total, 1{,}445 elements were labeled across the 84 requirements (\autoref{tab:goldstandard}).
The distribution is dominated by states (587), reflecting fine-grained use of that category, followed by activities (196), entities (119), components (88), data (59), nodes (35), data flows (35), and connections (33).
Despite interviewees being asked to reference specific architectural locations, 99 occurrences were classified as non-specific references; coreferences were used sparingly.

\newcommand*\rot[1]{\rotatebox{90}{\small #1}}
\begin{table}
\centering
\caption{Distribution of security elements in EVerest (n=1,445) over all 84 security requirements}
\label{tab:goldstandard}
\small
\begin{tabularx}{\columnwidth}{X *{9}{c} }
    \toprule
    & \rot{Components} & \rot{Data} & \rot{Nodes} & \rot{Entities} & \rot{States}  & \rot{Connections} & \rot{Data flows}  & \rot{Activities} & \rot{Control flows} \\
    \midrule
    named & 88 & 59 & 35 & 119 & 587 & 33 & 35 & 196 & 56 \\
    \rotatebox[origin=c]{180}{$\Lsh$} traced
           & 77 & 32 & -- & 3 & -- & -- & -- & -- & -- \\
    referenced & 96 & 33 & 4 & 16 & 5 & 2 & 15 & 10 & 12 \\
    coreferenced & 11 & 13 & 2 & 8 & 5 & 1 & 1 & 1 & 2 \\
    \midrule
    total & 195 & 105 & 41 & 143 & 597 & 36 & 51 & 207 & 70 \\
    \bottomrule
\end{tabularx}
\end{table}

Architectural trace links were annotated analogously, with conflicts resolved through inter-annotator agreement.
The gold standard provides model element IDs for entity-like elements (e.g., components, data, nodes, and entities) as visualized by the color-coded architecture in \autoref{fig:articts_overview}.
Overall, the labeling process required approximately 100 person-hours across three annotators.
\section{Summary and Outlook}

This paper presented the EVerest dataset, a multi-artifact resource based on EVerest, an industry-driven open-source software stack for \EV charging stations, spanning natural language documentation, security requirements, software architecture, and source code.
The 84 manually elicited security requirements are annotated with security objectives, 1{,}445 fine-grained security elements, acceptance windows, coreferences, and architectural trace links.
The dataset thereby addresses a recognized gap in the field, enabling research on security requirements classification, named entity recognition, architectural trace linking, and design-time as well as code-level security verification within a single resource.

During dataset creation, a concrete security weakness was identified using xDECAF~\cite{xDECAF}: Requirement~5 stipulates that \enquote{[\dots] tokens used for authentication should not be stored in plain text in log files or persistent storage [\dots]}.
The dataset snapshot's source code, however, violates this requirement~\footnote{Line~13 of auth\textunderscore token\textunderscore providerImpl.cpp in the PN532TokenProvider module.}.
The weakness (CWE-1295) was disclosed to PIONIX GmbH and remedied shortly thereafter, confirming the dataset's real-world relevance.

Future work will explore further weaknesses or vulnerabilities in the EVerest codebase and automated end-to-end security verification approaches leveraging the dataset's multi-artifact structure.

\begin{acks}
This work was funded by the Topic Engineering Secure Systems of the Helmholtz Association (HGF), supported by KASTEL Security Research Labs, Karlsruhe, by the pilot program Core Informatics at KIT (KiKIT) of the HGF, by the Deutsche Forschungsgemeinschaft (DFG) under the National Research Data Infrastructure – NFDI 52/1 – 501930651, and supported by the DFG - SFB 1608 - 501798263.
We thank Pionix, all EVerest developers, and our students for their contributions.
Generative AI tools were used for copy editing.
\end{acks}

\bibliographystyle{ACM-Reference-Format}
\bibliography{ase26}


\end{document}